\documentstyle[11pt,rotate,epsfig]{article}
\def\text{~}

\def\hc2{(\hbar c)^2}

\def\r2{\langle r^2 \rangle}
\def\Q2{$Q^2$}

\def\gev2c2{GeV$^2$/$c^2$}
\def\fm2{\text{fm}^2}
\def\a2         {{\mbox{$a_{2}                                        \
$}}}

\def\a2pigamma  {{\mbox{$a_2^-\rightarrow \pi^-\gamma                \
$}}}

\begin{document}
\pagestyle{empty}
 \begin{center}
  \vspace*{3.5cm}
 {\LARGE {\bf Color Transparency at COMPASS
via Exclusive Coherent Vector Meson Production
}\footnote{
 Talk presented
 at 2002 Praha Advanced Study Institute "Symmetries and Spin" Workshop, 
Praha-SPIN-2002, 
July 2002, Prague, Czech
Republic.  
The conference www 
site is 
http://mfinger.home.cern.ch/mfinger/praha2002//.}}\\

 \vspace*{1.8cm}

 \mbox{\Large 
   Murray Moinester$^{2} \!,$
 \,Oleg A. Grajek$^{3} \!$,
 \,Eli Piasetzky$^{2} \!$,
 \,Andrzej Sandacz$^{3}$}

 \vspace*{8mm}

 $^{2}$ School of Physics and Astronomy, R. and B. Sackler Faculty of
Exact
 Sciences,\\ Tel Aviv University, 69978 Ramat Aviv, Israel

 \vspace*{2mm}
 
 $^{3}$ So\l tan Institute for Nuclear Studies, ul. Ho\.{z}a 69, PL 00-681
Warsaw,
 Poland,
 
 \vspace*{4mm}

{\it E-mails: murraym@tauphy.tau.ac.il, 
oleg@fuw.edu.pl, eip@tauphy.tau.ac.il,
sandacz@fuw.edu.pl}
 \end{center}
 
 \vspace*{2cm}
 
 \begin{abstract}
 
 \vspace*{0.4cm}
 
 We examine the potential of the COMPASS experiment at CERN to study color
transparency via exclusive coherent vector meson production in hard
muon-nucleus scattering. It is demonstrated that COMPASS has high
sensitivity to test this important prediction of perturbative QCD.
 
 \vspace*{4.0cm}
 
 \end{abstract}

 \pagebreak
 
 \pagestyle{plain}
 \setcounter{page}{2}

 \section{Introduction}
 \label{lab_sec_1}
 
\hspace*{8mm}
One of predictions of perturbative QCD (pQCD)  is that at high $Q^2$ the
longitudinally polarized virtual photons $\gamma^{\ast}_{L}\/$ fluctuate
into hadronic components, e.g. $q\bar{q}$ pairs, whose transverse size $b =
\: \mid \! \bar{r}_{\! \perp q} - \bar{r}_{\! \perp \bar{q}} \! \mid $
decreases with $Q^2 \!$,   \,$b^2 \propto 
( 1/Q^{2} ).$ 
\,At large $Q^2$ the values of $b$ are
significantly smaller than the size of the nucleon. {\it Color
transparency\/} (CT)  is a phenomenon of pQCD, whose characteristic feature
is that small color-singlet objects interact with hadrons with small cross
sections \cite{low,bfgms,fms}.  Cross section for the interaction of such
small object, or small size configuration (SSC), with a hadron target has
been calculated in pQCD using the factorization theorem
\cite{bbfs,fms,frs}, as:
\begin{equation}
 \sigma _{q\bar{q}, \: N} = \frac{\pi^2}{3} \: b^2 \: \alpha_s(Q^2) \; x \;  
g(x, \: Q^2) \;, 
\end{equation} 
where $g(x, \: Q^2)$ is the gluon distribution function in the nucleon.

Observing CT in particular kinematics would prove experimentally the
applicability of the pQCD factorization theorem for those kinematics. Such a
proof provides important complementary support to a class of spin physics
experiments, for example related to measurements of generalized parton
distributions at COMPASS \cite {jp}.

The prerequisite for observing CT is to select a sample containing SSC's via
a hard process (i.e. with large $Q^{2}\!$, \,high $p^{}_{t}\/$, or large
produced mass).  For hard exclusive $\rho^0$ leptoproduction, in addition to
large $Q^2$, selection of the longitudinally polarized mesons is required.
To probe the SSC--nucleon cross sections one should study absorption of the
SSC propagating through nuclear matter. In order to clearly observe CT it is
necessary that the SSC lives long enough to traverse distances larger than
the size of the target nucleus, and that the SSC stays small while
propagating through the nucleus. These requirements are characterized in
terms of the coherence length $l_c$ (the distance traversed by the
$q\bar{q}$ fluctuation of the virtual photon in the target nucleon/nucleus
system)  and by the formation length (the distance in the target system
after a scattering needed for the $q\bar{q}$ fluctuation to develop into a
hadron).

  We consider here only 
exclusive {\it coherent\/} vector meson $V$ production at
small $|t|$. We a\-na\-ly\-sed the coherent and incoherent channels in Refs
\cite {asct,trentoct}. For near-forward ($t \approx 0$) coherent production and
complete CT, one expects that the cross section for the nuclear target is
related to that for the nucleon target, provided that $|t_{\rm min}< \!
R_A^{2} \! >/3| \ll 1 $, by:\\
\begin{equation} 
\frac{{\rm d}\sigma}{{\rm d}t}(\gamma ^* A \rightarrow V
A) \, |^{}_{t \approx 0} = A^2 \frac{{\rm d}\sigma}{{\rm d}t}(\gamma ^* N
\rightarrow V N) \, |^{}_{t \approx 0} \: . 
\end{equation}

Usually in experiments the $t$-integrated coherent cross section is
measured, for which CT predicts an approximate $A^{4/3}$ dependence.  
Without CT, if a hadron propagates in nuclear matter, the expected
$A$-dependence is weaker;  $A^{2/3}$ for the cross sections comparable to
the pion--nucleon cross section.

In searches for CT a commonly used quantity, measured in experiments, is the
{\it nuclear transparency\/} 
\begin{equation} 
T = \frac{\sigma^{}_{\! \!A}}{A \, \sigma^{}_{\! N}} \: , 
\end{equation} 
which is the ratio of the cross section per nucleon for a selected process
on a nucleus $A$ to the corresponding cross section on a free nucleon.

We concentrate on exclusive coherent vector meson production (EVMP), which could be
studied at the COMPASS experiment \cite{compass,bradamante}. Kinematic
variables used in this paper are listed in Table \ref{kinvar}.

%%%%%%%%%%%%%%%%%%%%%%%%%%%%%%%%%%%%%%%%%%%%%%%%%%%%%%%%%

\begin{table}[t]
\caption{Kinematic variables used in the text.
\label{kinvar}}
\vspace{0.2cm}
\begin{center}
\footnotesize
\begin{tabular}{|l|l|}
\hline
\hline
$k$, $k'$         &   four-momentum of the incident and scattered
muon,\\
$p$, $v$          &   four-momentum of the target nucleon and vector
meson $V\!$,\\
$q = k - k^{\prime}$  &  four-momentum of the virtual photon,                   \\
$Q^2 = -q^2$ &  virtuality of the photon
$\gamma^{\ast} \!$,           \\
$\nu =(p\cdot q)/M^{}_{\! p}$ & energy of the virtual photon in the
laboratory system,\\
    & $M^{}_{\! p}$ is the proton mass,\\       
$x = Q^2/(2 M^{}_{\! p} \nu )$ & Bjorken scaling variable,\\
$t = (q-v)^2$ &
squared four-momentum transfer to the target, \\
$p_t^2$      & transverse-momentum squared of the vector meson, \\
$m^{}_{\! V} = (v^2)^{\frac{1}{2}}$ & invariant mass of the vector meson
$V\!$,\\ 
$M^{2}_{\! X} = (p+q-v)^2$ & missing-mass squared of the undetected
recoiled system,\\
$I = (M^{2}_{\! X} - M^{2}_{\! p})/W^2$ & inelasticity \\
\hline
\hline
\end{tabular}
\end{center}
\end{table}

In order to study CT for the exclusive production of mesons composed of
light quarks (like $\rho$) one should select {\it large\/} $Q^2 \!$, \,{\it
moderately small\/} $x$ and {\it longitudinally\/} polarized mesons.
 
The $t$-dependence of the cross section for exclusive meson production on
the nucleon is approximately exponential, $e^{B(t-t_{\min})}$, where
$t_{\rm min} \approx
-M_p^2x^2(1+ m_{\! V}^2/Q^2)^2$ 
and $m_{\! V}$
is the mass of the produced meson. For hard EVMP, if SSC's are important,
the slope $B$ significantly decreases with increasing $Q^2\!$, \,approaching
a universal value ($4.5-5\:$ GeV$^{-2}$) related to the nucleon radius. This
is so because the transverse interquark distance in the SSC decreases with
increasing $Q^2$.

For EVMP on a nucleus $A$ the $t$-dependence of the cross section
is approximately reproduced by a sum of two exponential functions.  The peak
at the lowest $|t|$ values, with the slope proportional to the
nucleus radius squared, $< \! R^{2}_{\! A} \! >$, is mostly due to the
coherent production, whereas at somewhat larger $|t|$ the incoherent
production on quasi-free nucleons dominates and the slope is almost equal to
that for the production on free nucleons. Applying cuts on $t$
allows to select successfully samples of events, which are strongly
dominated by either coherent or incoherent production, as discussed in Refs.
\cite{NMCJpsi,Sokolov,Wei97}.

Strong recent evidence for CT comes from Fermilab E791 experiment on the
\mbox{$A$-de-} pendence of coherent diffractive dissociation of pions into
two high-$p^{}_{t}\/$ jets \cite{Wei97}. Also the E691 experiment results on
$A$-dependence of coherent $J \! / \! \psi\/$ photoproduction \cite{Sokolov}
and the NMC measurements of $A$-dependence of coherent and quasielastic $J
\! / \! \psi\/$ muoproduction are {\it consistent\/} with CT, but don't
demonstrate it unambiguously.

\section{Experimental method}

\label{lab_sec_2}

\hspace*{8mm}
We proposed \cite {asct} to study CT via {\it exclusive
coherent $\rho$ meson production\/} $\mu A \rightarrow \mu \, \rho A$ on C
and Pb nuclei, as well as the exclusive production of other vector mesons.  
The preferable decay modes are those into the charged particles
such as $\rho ^0 \rightarrow
\pi^{+} \pi^{-} \!$.

We propose to complement the initial setup of the COMPASS \cite{compass,bradamante}
for the muon-beam run with the polarized target by adding the nuclear
targets downstream of the polarized target.

The high-intensity high-energy incident muon-beam will impinge on the
polarized target and a downstream thin nuclear target. The momenta of the
scattered muon and of the produced charged particles will be reconstructed
in the two magnetic spectrometers, using the magnets SM1 and SM2,
instrumented with various tracking detectors.  In order to select exclusive
events, we rely only on the kinematics of the scattered muon and the
produced meson.

The trigger will use the information from hodoscopes registering the
scattered muons, and from calorimeters registering deposits of energy of the
particles in the final state.

\begin{figure}[t]

\begin{center}
\mbox{\epsfig{file=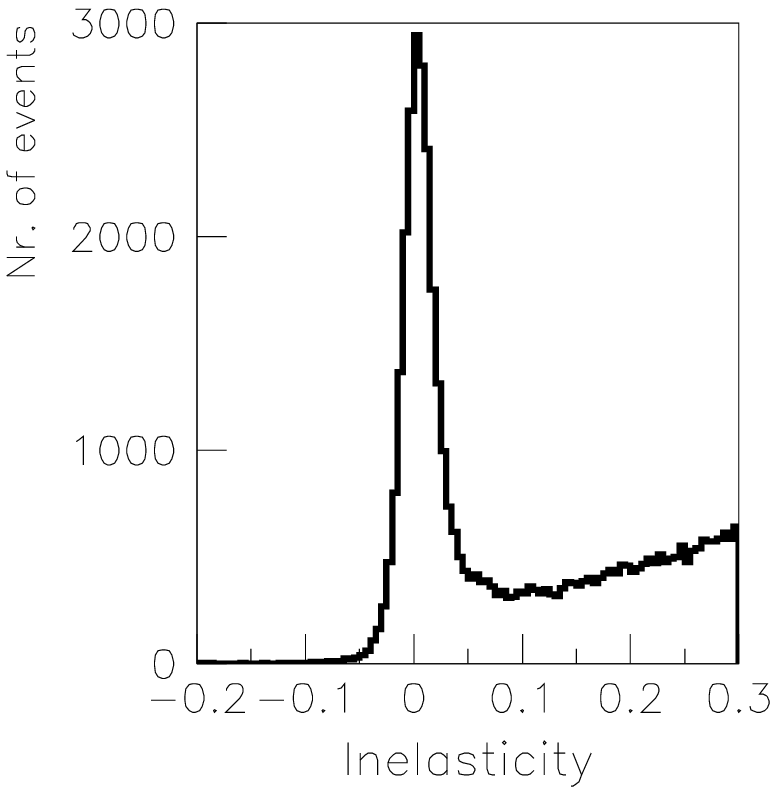,height=3.5in}} 
\end{center}

\caption{Inelasticity distribution. The SMC preliminary results \cite
{smcrhoq} for $\mu N \rightarrow \mu \rho ^0 N$.
\label{ange}}
\end{figure}

The discrimination of non-exclusive events will be done by applying cuts on
the inelasticity $I\/$ (for the definition see Table \ref{kinvar}). In Fig.
\ref{ange} the inelasticity distribution is shown for the SMC $\rho ^0$
sample ~\cite{smcrhoq} for the events with the invariant mass in the central
part of the $\rho ^0$ peak. For the inelasticity distribution the peak at
$I=0$ is the signal of exclusive $\rho ^0$ production. Non-exclusive events,
where in addition to detected fast hadrons, slow undetected hadrons were
produced, appear at $I>0$.  
With larger statistic in COMPASS it will be
possible to apply more tight inelasticity cuts in order to 
further reduce the
background.

The selections of coherent or incoherent production will be done on a
statistical basis, using the $t$-distribution; at the lowest
$|t|$ values coherent events predominate.

Separation of the $\rho^{0}$ samples with the enhanced content of
longitudinally or transversely polarized mesons will be done by applying
cuts on the measured angular distributions of pions from the decays of the
parent $\rho^{0}$.

The minimal covered $Q^2$-range is expected to be $0.05 < Q^2 < 10
\:\rm{GeV}^2 \!$. \,For the \mbox{med-} \linebreak ium and large $Q^2$
values $(Q^2 > 2\:\rm{GeV}^2)$ the range $0.006 < x < 0.1$ will be covered
with good acceptance.

The basic observable for each process studied will be the ratio of the
nuclear transparencies for lead and carbon, $R^{}_{\rm T} =
T_{\rm{Pb}}/T_{\rm{C}} = (\sigma _{\rm{Pb}}/A_{\rm{Pb}})/(\sigma
_{\rm{C}}/A_{\rm{C}})$.

\section{Simulation of exclusive $\rho^{0}$ events}

\label{lab_sec_3}

\hspace*{8mm} We present details of the simulation of exclusive coherent
$\rho ^0$ production $(\mu A \rightarrow \mu \, \rho^{0} A)$ in the COMPASS
experiment with the carbon and lead targets. The simulations were done with
a dedicated fast Monte Carlo program which generates deep inelastic
exclusive $\rho^{0}$ events with subsequent decay $\rho ^0 \rightarrow
\pi^{+} \pi^{-} \!$. \,At this stage there was no attempt to include any
background in the event generators.  
Details of the used parameterization of
the cross section for the production of $\rho^{0}$ on the free nucleon, $\mu
\, N \rightarrow \mu \, \rho^{0} N$, with the subsequent decay $\rho^{0}
\rightarrow \pi^{+} \pi^{-}$ were presented in Ref. \cite{asct}.
We relate the differential cross sections for the free nucleon
to these for coherent production on the nucleus $A$ via
\begin{equation}
\mbox{\Huge (} \frac{{\rm{d}} \sigma^{\rm coh}_{\! \! A}}{{\rm{d}} t}
  \mbox{\Huge )}_{\! \! i} =
  A^{2}_{\rm eff \!, \; coh} \cdot e^{< R^{2}_{\! A} > \, t/3} \cdot
\mbox{\Huge (} \frac{{\rm{d}} \sigma^{}_{\! N}}{{\rm{d}} t} \mbox{\Huge
  )}^{}_{\! \! i} \;\: .
\label{EQ_MC50}
\end{equation}
\noindent Here $< \! R^{2}_{\! A} \! >$ is the mean squared radius of the
nucleus, $A^{}_{\rm eff \!, \; coh}$ takes account of nuclear screening for
the coherent process, and $i = L\/$ or $T\/$ designates the polarization of
$\rho^{0}\!$. \,We used the approximation $t - t_{min} \simeq -p^{2}_{t}\/$.

We generated the cross sections for two models. For the complete color
transparency model (CT model)  we used $A^{}_{\rm eff \!, \; coh} = A\/$.  
In another model we assumed a substantial nuclear absorption (NA model) and
used $A^{}_{\rm eff \!, \; coh} = A^{0.75 \!}$.

Details of the simulation of the apparatus effects were given in \cite{asct}. Here we discuss only the
trigger simulation.  To simulate the trigger acceptance a trajectory of the
scattered muon behind the second magnet was calculated, and the hits in the
muon hodoscopes were checked.  We consider two different trigger
acceptances, the Medium $Q^{2}$ range Trigger (MT) and the Full $Q^2$ range
Trigger (FT). The $Q^2$-dependence of the trigger efficiency is denoted
$\epsilon^{}_{\rm tr}\/$. For the MT trigger at 190 GeV beam,
$\epsilon^{}_{\rm tr}$ decreases quickly with $Q^2$ from 0.7 at $Q^2 = 2
\:\rm{GeV}^2$ to about 0.1 at $Q^2 = 10 \:\rm{GeV}^2 \!$. \,For the FT
trigger the $Q^2$-dependence is weaker and the trigger acceptance is higher;
it is always bigger than 0.5 for $Q^2 < 70 \:\rm{GeV}^2$ at 190 GeV beam
energy.

\section{Results on exclusive $\rho ^0$ production}

\label{lab_sec_4}

\hspace*{8mm} We considered 190 GeV muon beam. The simulations were done
independently for the carbon $(A = 12)$ and lead targets $(A = 207)$, and
for two triggers (MT and FT). For each target and each trigger we assumed
two different models describing the nuclear effects for exclusive $\rho^{0}$
production: CT model and NA model (cf. Section 3).

The kinematic range considered was the following:
\begin{equation}
\label{EQ_MC5} 2 < Q^2 < 80\: \rm{GeV}^2 \: ,
\end{equation}
\begin{equation}
\label{EQ_MC6}
35 < \nu < 170\: \rm{GeV} \: .
\end{equation}
The upper cut on $\nu $ was chosen to eliminate the kinematic region where
the amount of radiative events is large, whereas the lower one to eliminate
the region where the acceptance for pions from $\rho^{0}$ decay is low.

The total efficiency $\epsilon^{}_{\rm tot}\/$ to observe exclusive
$\rho^{0}$ events results from: the acceptance of the trigger
$(\epsilon^{}_{\rm tr})\/$, acceptance to detect the pions
$(\epsilon^{}_{\rm had})\/$, efficiency for tracks reconstruction
$(\epsilon^{}_{\rm rec})\/$, cut on the invariant mass of two pions, $0.62 <
M^{}_{\pi \pi } < 0.92 \; \rm{GeV} \!$, \,used for the selection of the
samples ($\epsilon^{}_{\rm mass}\/$), and efficiency for an event to survive
the secondary interactions $(\epsilon^{}_{\rm sec})\/$. The contributions of
all these effects to $\epsilon^{}_{\rm tot}\/$ are similar to those
presented in \cite{memo2000} for the polarized target, except for the
effects of the secondary interactions. The approximate value of
$\epsilon^{}_{\rm sec}\/$ is equal to 0.87 for the carbon target and 0.94
for the lead target. The total efficiency $\epsilon^{}_{\rm tot}\/$ is about
0.48 for the FT trigger and about 0.30 for the MT trigger.

The total expected cross section for exclusive $\rho ^0$ production on the
nucleon is
\begin{equation}
 \sigma^{\rm tot}_{\mu N \rightarrow \mu \rho^{0} N} = \int_{\nu^{}_{\rm
min}}^{\nu^{}_{\rm max}}
  \! \! \int_{Q^{2}_{\rm min}}^{Q^{2}_{\rm max}}
  \! \! \sigma^{}_{\mu N \rightarrow \mu \rho^{0} N} (Q^{2} \!, \: \nu )
\:
  {\rm d} Q^{\! 2} \, {\rm d} \nu \; , 
\end{equation}
where the kinematic range was defined before. The value of $\sigma^{\rm
tot}_{\mu N \rightarrow \mu \rho^{0} N}\/$ is 283 pb. For nuclear targets
the corresponding values depend on $A$ and on the assumed model for nuclear
absorption.

The expected muon beam intensity will be about $10^{8} \! / \! \rm{s}$
during spills of length of about 2$\:$s, which will repeat every 14.4$\:$s.
With the proposed thin nuclear targets, each of 17.6 g/cm$^2$, the
luminosity will be ${\cal{L}} = 12.6 \:\rm{pb}^{-1} \cdot \rm{day}^{-1} \!$.
\,The estimates of the numbers of accepted events were done for a period of
data taking of 150 days (1 year), 
 divided equally between two targets. An overall SPS and
COMPASS efficiency of 25\% was assumed. The numbers of accepted events for
the carbon and lead targets, assuming the two models for the nuclear
absorption mentioned earlier, are given in Table \ref{rates}.

%%%%%%%%%%%%%%%%%%%%%%%%%%%%%%%%%%%%%%%%%%%%%%%%%%%%%%%%%

\begin{table}[t]

\vspace*{-5mm}

\caption{Numbers of accepted events for two considered models
	 of the nuclear absorption.
\label{rates}}
%\vspace{0.2cm}
\begin{center}
\footnotesize
\begin{tabular}{|c|c|c|}
\hline
\hline
  \raisebox{0mm}[5mm][3mm]{\hspace*{3mm} {\large model} \hspace*{3mm}}  &
{\large \hspace*{6mm} $N^{}_{\rm C}$ \hspace*{6mm}}
   &  {\large \hspace*{6mm} $N^{}_{\rm Pb}$ \hspace*{6mm}}  \\
\hline
\hline
  \raisebox{0mm}[5mm][3mm]{\large CT}   &  {\large 70 000}  &  {\large 200
000}  \\
\hline
  \raisebox{0mm}[5mm][3mm]{\large NA}   &  {\large 28 000}  &  {\large 20
000}   \\
\hline
\hline
\end{tabular}
\end{center}
\end{table}

%%%%%%%%%%%%%%%%%%%%%%%%%%%%%%%%%%%%%%%%%%%%%%%%%%%%%%%%%

The distributions of accepted exclusive events as a function of $x$, $Q^2$
and $\nu $ are similar to those presented in \cite{memo2000}.

\begin{figure}[t]
 
\begin{center}
\mbox{\epsfig{file=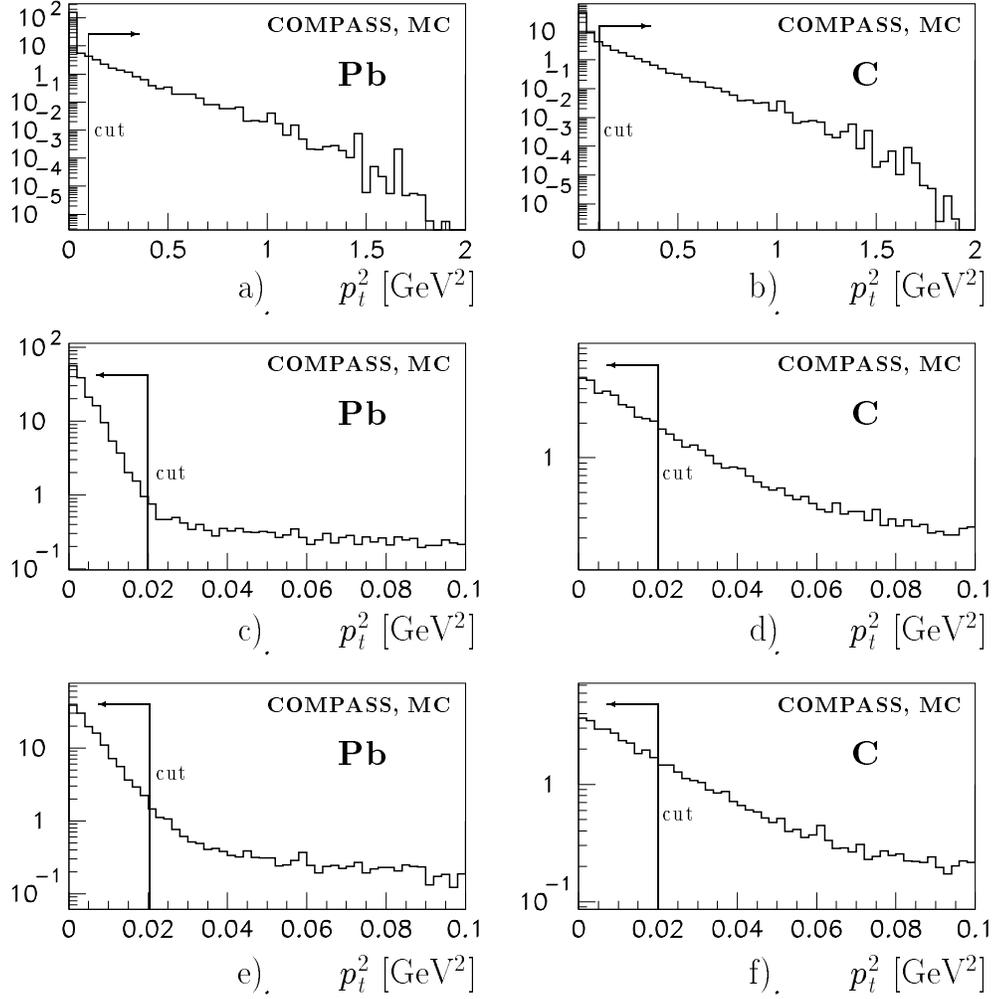,height=15cm}}
\end{center}

\caption{Distributions of $p^{2}_{t}\/$ for the lead (left) and carbon
(right) targets. The distributions a) and b) at the top are for the
generated events and the wide range of $p^{2}_{t}\/$, whereas those in the
middle, c) and d), correspond to the range of low $p^{2}_{t}\/$. The bottom
distributions, e) and f), are for the accepted events, with the kinematical
smearing taken into account. The arrows show the cuts described in the text.
\label{pt2}}
\end{figure}

In Fig.$\;$\ref{pt2} we present the $p^{2}_{t}\/$ distributions for both
targets. We observe clear coherent peaks at small $p^{2}_{t}\/$ ($< 0.05
\:\rm{GeV}^2$) and less steep distributions for the incoherent events at
larger $p^{2}_{t}\/$. The arrows at the top histograms indicate the cut
$p^{2}_{t} > 0.1 \:\rm{GeV}^2 \!$, \,used to select the incoherent samples.  
For the middle and bottom histograms the arrows indicate the cut $p^{2}_{t}
< 0.02 \:\rm{GeV}^2 \!$, \,used to select the samples which are dominated by
coherent events --- the \linebreak so called coherent samples. For the
samples defined by the latter cut the fraction of the incoherent events is
at the level of up to 10\%, depending on the nucleus and on the model for
nuclear absorption.

The effect of the kinematical smearing on $p^{2}_{t}\/$ may be seen by
comparing the distributions for the generated events (middle row) to the
ones for measured events (bottom row) where the acceptance and smearing were
included. The smearing of $p^{2}_{t}\/$ increases with increasing
$p^{2}_{t}\/$; it is about $0.006 \:\rm{GeV}^2$ for the coherent samples
($p^{2}_{t} < 0.02 \:\rm{GeV}^2$) and about $0.03 \:\rm{GeV}^2$ for the
incoherent samples ($p^{2}_{t} > 0.1 \:\rm{GeV}^2$).

\begin{figure}[t]
\begin{center}
\mbox{\epsfig{file=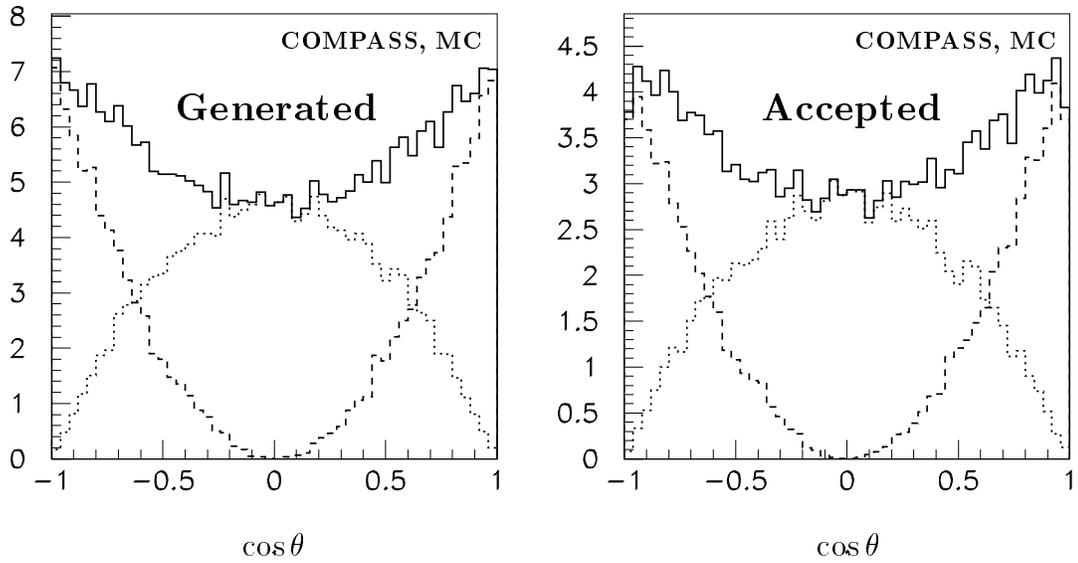,height=10cm}}
\end{center}

\caption{Distributions of $\cos \theta \/$ for the pions $(\pi^{+})$ from
the decays of $\rho^{0}\/$ mesons produced on lead, for generated (left) and
accepted (right) events. For the latter ones the effects of the kinematical
smearing were taken into account. The dashed- and dotted-line histograms are
for the longitudinally and transversely polarized $\rho^{0}\/$'s,
respectively, and the solid-line histograms are for their sum.
\label{costh}}
\end{figure}

The analysis of the $\rho^{0}$ decay distributions allows us to study
spin-dependent properties of the production process \cite{SW}, in particular
the polarization of $\rho^{0}\!$. 
The $\rho^{0}$ decay angular distribution
$W(\cos \theta, \: \phi )$
is studied in the $s$-channel helicity frame.
\,The $\rho^0$ direction in the virtual
photon--nucleon centre-of-mass system is taken as the quantization axis. The
angle $\theta \/$ is the polar angle and $\phi\/$ the azimuthal angle of the
$\pi^{+}\/$ in the $\rho^{0}$ rest frame. The $\cos \theta \/$ distributions
for pions from $\rho^{0}$ decays are shown in Fig.$\;$\ref{costh} for the
lead target.  The distributions for longitudinally (dashed lines) and
transversely (dotted lines) polarized parent $\rho^{0}$'s are markedly
different.  Their sum is also indicated.

As $Q^2$ increases, the approach to the CT limit is expected to be different
for EVMP by longitudinally polarized virtual photons from that by
transversely polarized photons. To increase the sensitivity of the search
for CT, we study $A$-dependence of the cross sections for samples with
different $\rho ^0$ polarizations, which will be selected by cuts on $\cos
\theta \/$. Such cuts will allow us to select the samples with enhanced
contributions of the events initiated by the virtual photons of a desired
polarization. Studies of the samples with {\it different polarizations\/} of
the virtual photons are {\it necessary\/} for the unambiguous demonstration
of CT.

Another aspect which is important for CT studies, is the covered range of
the coherence length $l^{}_{c}\/$ (cf. Section 1).  We showed in Ref. \cite{asct}
 the
correlation of $l^{}_{c}\/$ vs. $Q^{2}\/$ for a sample of accepted events.
At small $l_c$ values
the effects of initial and final state interactions in the nuclei mimic CT
\cite{HERMES}. It is important to disentangle effects
due to CT from those caused by the modified absorption at small $l_c$
values.

For a limited statistics, we may use the combined data at $l_c$ values
exceeding the sizes of the target nuclei. The radius of the carbon nucleus
is $< r^{2}_{\rm C} >^{1/2} = 2.5\:\rm{fm}$ and that of the lead nucleus is
$< r^{2}_{\rm Pb} >^{1/2} = 5.5\:\rm{fm}$. Therefore, one may use the
selection $l^{}_{c} > l^{\rm min}_{c} \simeq 2 \cdot < r^{2}_{\rm Pb}
>^{1/2} = 11 \:\rm{fm}$. About a half of events survive such a cut on
$l^{}_{c}$. These events cover the range of $Q^{2} < 6 \:\rm{GeV}^2 \!$,
\,which is expected to be sufficient to observe CT.

\begin{figure}[t]
\begin{center}
\mbox{\epsfig{file=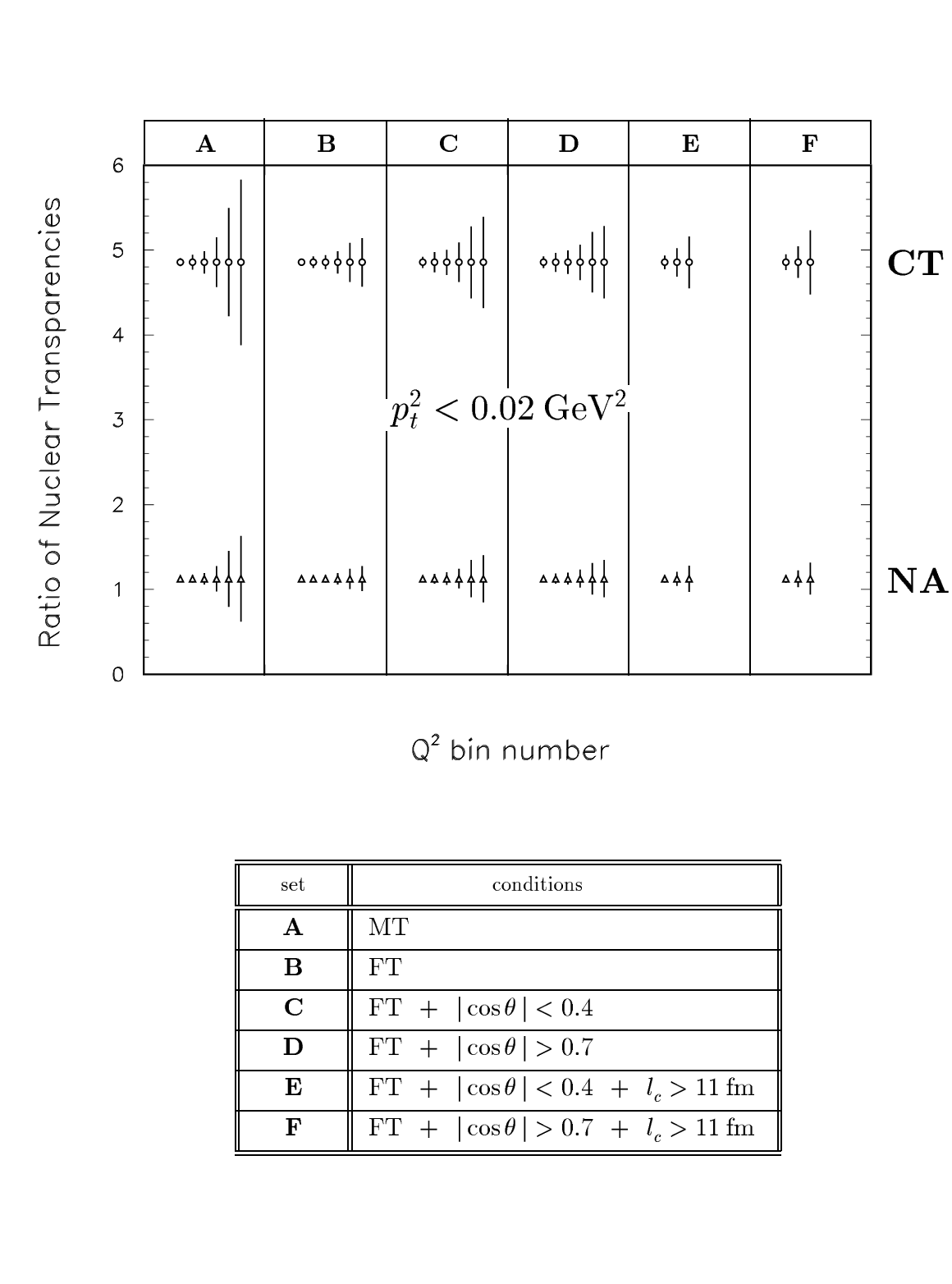,height=18cm}}
\end{center} 

\caption{Expected statistical precision of the measurement of the ratio
$R^{}_{\rm{T}}\/$ of $\:\rm{GeV}^2$ nuclear transparencies for lead and
carbon for $p^{2}_{t} < 0.02 \:\rm{GeV}^2$ and for different $Q^{2}$ bins,
for CT (upper band) and for NA (lower band) models. The $Q^{2}$ bins are
defined in Table 3. Sets A, B, C, D, E and F correspond to the conditions
specified in the table below the plot.
\label{ratcoh}}
\end{figure}

The estimated values and statistical precision of $R^{}_{\rm T}$, the ratio
of the nuclear transparencies for lead and carbon, are presented for
different $Q^{2}$ bins in Fig.$\;$\ref{ratcoh} for $p^{2}_{t} < 0.02
\:\rm{GeV}^{2}$. \,The $Q^{2}$ bins are specified in Table \ref{q2bins}.
Figure$\;$\ref{ratcoh} comprises predictions for two models, CT and NA, and
for 6 different samples of accepted events for each model. For each sample a
set of "measurements" in different $Q^2$ bins is shown. Sets {\bf A} and
{\bf B} were obtained using the standard selections for the MT and FT
triggers, respectively. Four remaining sets were obtained for the FT trigger
with additional selections: {\bf C} with $\mid \! \cos \theta \! \mid <
0.4$, {\bf D} with $\mid \! \cos \theta \! \mid > 0.7$, {\bf E} with $\mid
\! \cos \theta \! \mid < 0.4$ and $l^{}_{c} > 11\:\rm{fm}$, and {\bf F} with
$\mid \! \cos \theta \! \mid > 0.7$ and $l^{}_{c} > 11 \:\rm{fm}$. Note that
for sets {\bf E} and {\bf F} only three lower $Q^{2}$ bins appear. One
expects large differences in $R^{}_{\rm T}$ for the two considered models.
For coherent samples $R^{}_{\rm T} \approx 5$ for CT model and $\approx 1$
for NA model. At $Q^2 \simeq 5 \:\rm{GeV}^2$ the precision of the
measurement of %$r_T$ for coherent events will be better than 17\%, even for
$R^{}_{\rm T}$ for coherent events will be better than 17\%,
even for the restricted samples {\bf E} and {\bf F}, thus allowing excellent
discrimination between the two models for the nuclear absorption.

%%%%%%%%%%%%%%%%%%%%%%%%%%%%%%%%%%%%%%%%%%%%%%%%%%%%%%%%%

\begin{table}[t]

\vspace*{-5mm}

%\caption{$Q^{2}$-bins used for the determination of $r_T$.
\caption{$Q^{2}$ bins used for the determination of $R^{}_{\rm T}$.
\label{q2bins}}
%\vspace{0.2cm}
\begin{center}
%\footnotesize
\begin{large}
\begin{tabular}{|c|c|c|c|}
\hline
\hline
 \raisebox{0mm}[5mm][2mm]{Bin number}  &  \ \ \ \ $Q^{2}$ bin\ \ \ \   &
  			  \ \ \ \ $<Q^{2}>$\ \ \ \   &  \ \ \ \ $<x>$\ \ \
\   \\
   & \raisebox{0mm}[4mm][3mm]{$[\rm{GeV}^2]$} & $[\rm{GeV}^2]$ &\\
\hline
\hline
  \raisebox{0mm}[5mm][3mm]{1}  &  2--3  &  2.4  &  0.016  \\
\hline
  \raisebox{0mm}[5mm][3mm]{2}  &  3--4  &  3.4  &  0.022  \\
\hline
  \raisebox{0mm}[5mm][3mm]{3}  &  4--6  &  4.8  &  0.031  \\
\hline
  \raisebox{0mm}[5mm][3mm]{4}  &  6--9  &  7.2  &  0.048  \\
\hline
  \raisebox{0mm}[5mm][3mm]{5}  &  9--12  &  10.2  &  0.072  \\
\hline
  \raisebox{0mm}[5mm][3mm]{6}  &  12--20  &  14.8  &  0.11  \\
\hline
\hline
\end{tabular}
\end{large}
\end{center}
\end{table}

%%%%%%%%%%%%%%%%%%%%%%%%%%%%%%%%%%%%%%%%%%%%%%%%%%%%%%%%%

\section{Comparison with previous experiments}

\label{lab_sec_5}

\hspace*{8mm}We compare COMPASS capabilities to demonstrate CT to those of
previous experiments in which exclusive coherent $\rho ^0$ leptoproduction
on nuclear targets was studied.  For the comparison we must also select
experiments which covered $Q^2$ range extending to large values, bigger than
2 GeV$^2$.  The only experiment which published such high $Q^2$ exclusive
coherent data is the NMC experiment \cite{nmc}.

NMC has published data on coherent exclusive $\rho ^0$ production on $^2{\rm
H}$, C and Ca targets. The muon beam energy was 200 GeV and the data cover
the ranges $2 < Q^2 < 25\: {\rm GeV}^2$ and $1 < l_c < 30$ fm. Due to the
moderate statistics of the data the $Q^2$-dependence of nuclear absorption
was not obtained at sufficiently large or at fixed $l_c$, which is necessary to
unambigously demonstrate CT.

The COMPASS data at medium and large $Q^2$ (FT trigger) will cover the
kinematic range similar to that of the NMC data. The expected statistics for
the carbon target will be about 2 orders of magnitude higher than that of
the NMC data for the same target. This increase is in particular due to
higher beam intensity and larger acceptance in the COMPASS experiment.
In addition COMPASS will use also lead target and will extend measurements to small
$Q^2$. 
Due to large statistics splitting of COMPASS data in
several $Q^2$ and $l_c$ bins as well as the selection of events with
longitudinal or transverese $\rho ^0$ polarization will be possible.

\section{Conclusions}

\label{lab_sec_7}

\hspace*{8mm}We have simulated and analysed exclusive
$\rho^{0}$ muoproduction at
the COMPASS experiment using thin nuclear targets of carbon and lead. For
muon beam energy of 190 GeV and a trigger for medium and large $Q^{2}$, the
covered kinematic range is $2 < Q^2 < 20\: \rm{GeV}^2$ and $35 < \nu < 170\:
\rm{GeV}$.

Good resolutions in $Q^{2} \!$, \,$l^{}_{c}$, $t$ ($p^{2}_{t}$) and
$\cos \theta \/$ are feasible. An efficient selection of coherent events is
possible by applying cuts on $p^{2}_{t}$. In order to obtain the samples of
events initiated with a probability of about 80\% by either
$\gamma^{\ast}_{L}\/$ or $\gamma^{\ast}_{T}\/$, the cuts on the $\rho^{0}$
decay angular distribution of $\cos \theta \/$ will be used. The search for
CT could be facilitated by using the events with $l^{}_{c}\/$ values
exceeding the sizes of the target nuclei. The fraction of such events is
substantial and the covered $Q^{2}$ range seems sufficient to observe CT.

We showed high sensitivity of the measured ratio $R^{}_{\rm T}$ of nuclear
transparencies for lead and carbon for different models of nuclear
absorption. Good statistical accuracy of the measured $R^{}_{\rm T}$ may be
achieved already for one year of data taking. These measurements, taken at
different $Q^{2}$ intervals, may allow to discriminate between different
mechanisms of the interaction of the hadronic components of the virtual
photon with the nucleus.

In conclusion, the planned comprehensive studies of exclusive vector meson
production on different nuclear targets at the COMPASS experiment are able
to unambiguously demonstrate CT.

%\newpage
\section{Acknowledgements}
\label{lab_sec_8}

\hspace*{8mm}The authors gratefully acknowledge useful discussions with L.
Frankfurt. The work was supported in part by the Israel Science Foundation
(ISF) founded by the Israel Academy of Sciences and Humanities, Jerusalem,
Israel and by the Polish State Committee for Scientific Research (KBN SPUB
Nr 621/E-78/SPUB-M/CERN/P-03/DZ 298/2000 and KBN grant Nr 2 P03B 113 19).
One of us (A.S.) acknowledges support from the Raymond and Beverly Sackler
Visiting Chair in Exact Sciences during his stay at Tel Aviv University.

%-----------------------------------------------------------------------\\

%  FIGURES
%  =======

\newpage

\newpage

\newpage

\newpage

\end{document}